# On the Dynamics of Light Quarks in QCD

Khalil Bitar, Robert G. Edwards*, Urs M. Heller, and A. D. Kennedy
Supercomputer Computations Research Institute
Florida State University
Tallahassee, FL 32306-4052, USA
Internet: {kmb,edwards,heller,adk}@scri.fsu.edu

We describe recent results concerning the behavior of lattice QCD with light dynamical Wilson and Staggered quarks. We show that it is possible to reach regions of parameter space with light pions $m_\pi \approx 0.2/a$ using Wilson fermions. If the Hybrid Molecular Dynamics (HMD) algorithm is used with the same parameters it gives incorrect results. We also present preliminary results using a higher-order integration scheme.

## 1. Introduction

In this paper, we set out with the objective of examining the effects of $\delta H$ errors in the integration of Hamilton's equations of motion. We have performed tests using 2 flavors of Wilson fermions at $\beta = 5.30, \kappa = 0.1675$ and $0.1677$ on a $16^3 \times 32$ lattice. We have also studied four flavors of Staggered fermions at $\beta = 5.05$, $m_q = 0.005$ on an $8^3 \times 16$ lattice. Previous studies of step-size errors including fermions were limited to relatively small lattices and large pion masses. Our goal is to extend these studies to parameter regions that will be realistic on future machines.

The standard methods of simulating QCD with dynamical fermions are Hybrid Monte Carlo (HMC) via the $\phi$-algorithm [1] and Hybrid Molecular Dynamics (HMD) via the $R$-algorithm [2]. The major distinction between the two algorithms is how the fermionic determinant is simulated and how the configuration obtained by approximately integrating Hamilton's equations of motion is handled. The nomenclature is somewhat confusing. The expression Hybrid Molecular Dynamics is principally used to name the $R$-algorithm; however, it is also used to describe the $\phi$-algorithm without the acceptance test. The principle advantage of the $\phi$-algorithm with the acceptance test (HMC) is that the algorithm is exact – no finite step-size errors are introduced into the equilibrium probability distribution.

Our tests are limited to the Hybrid Monte Carlo algorithm where we can compare the algorithm with and without the Metropolis acceptance. We will call the algorithm without the Metropolis acceptance the HMD algorithm. All our computations were done on the Connection Machine CM-2 at SCRI using a conjugate gradient linear equation solver (written in CMIS) which runs at $> 5$ Gigaflops.

A major difficulty in simulating QCD is the long exponential autocorrelation times for the system to decay into equilibrium and the long integrated autocorrelation times while in equilibrium. In fact, our results for Wilson fermions with the lightest pion ($m_\pi \approx 0.2/a$) indicate a decay time of about 700 unit-length trajectories. We *should* expect this behavior. The pion is the lightest excitation of QCD, because it is trying to be a Goldstone boson. Unlike the quenched approximation where the lightest dynamical excitation is a glueball, for full QCD the pion mass is the inverse correlation length, $m_\pi = 1/\xi$. Thus, for $m_\pi = 0.2/a$ we expect that the correlation length $\xi \approx 5a$ in equilibrium; it is only reasonable to expect an algorithm to take much longer to develop such long distance correlations than it takes to equilibrate a quenched system with $\xi \approx a$. Theoretically, we expect the characteristic relaxation time to be the same as the autocorrelation time, which in turn is proportional to $\xi^z$ where $z$ is the dynamical critical exponent. For HMC it is believed that $1 \le z \le 2$ [5].

How can we measure the correlation length of an ensemble of configurations? The obvious method of measuring $\xi$ is to measure the pion mass by measuring the asymptotic decay of a pion correlation function $\langle \pi(0)\pi(t) \rangle$. If we have a set of equilibrated configurations this is a completely valid procedure, but consider what would happen if the Markov process had not yet converged to the true equilibrium distribution. Because the system starts from a configuration which does not have the true long distance correlations built in, the unequilibrated configurations will tend to have $\xi < 1/m_\pi$. Roughly speaking we can say that they correspond more closely to the equilibrium distribution of QCD with a larger value of the quark mass $m_q$– we shall call this the "sea" quark mass – than the actual value appearing in the action. Of course, the actual distribution of some set of unequilibrated configurations is some complicated mess, but it is certainly reasonable to assume that such a shift in $m_q$ is the dominant effect when near $m_{crit} = 0$ for Staggered fermions or $\kappa_c$ in the case of Wilson fermions.

---

*Speaker at the conference

The quark mass does not only appear in the action, it is also explicitly present in the pion operator $\pi$ used to measure the pion mass. Since this quark mass has nothing to do with the dynamics of the system we shall call it the "valence" quark mass. For our computations we expect that the sea $m_q$ approaches the valence $m_q$ from above as the system approaches equilibrium: in other words the correlation length estimated using the correlation function for a "valence" pion will be larger than the actual correlation length of the system until the system reaches equilibrium.

For a quenched computation or for an inexact dynamical quark algorithm valence and sea parameters are different even in equilibrium. One illustration of these ideas is that it is easy to measure a light pion mass on a quenched configuration, even though the correlation length is much smaller than the inverse pion mass.

The existence of a light valence pion is not evidence for a system containing light dynamical pions unless the system can be shown to have equilibrated to the correct probability distribution. A good indicator we have found is blocked gauge field operators (using Teper blocking), for example the blocked spatial plaquette. Blocked plaquettes do not need an extra parameter and are sensitive to the longer range properties of the system. The effects of dynamical quarks should appear as a modification of the quenched QCD beta function at longer distances. The blocked observables indicate coherence and correlation in the system.

The proceeding general arguments are applicable to both Wilson and Staggered fermions. The difference between the two fermions is at which quark mass scale should the step-size errors significantly effect the equilibrium probability distribution. The fermionic contribution to the molecular dynamics force is

$$\chi \left(M^\dagger M\right)^{-1} \frac{\delta}{\delta U} \left[M^\dagger M\right] \left(M^\dagger M\right)^{-1} \chi \qquad (1)$$

where $M = M(U(t))$, $U(t)$ is the gauge field configuration at integration time $t$ and $\chi$ is the pseudo-fermion field. When the quark mass approaches criticality the fermionic matrix becomes singular and the smallest eigenvalue dominates the force. The lowest eigenvalue $\lambda_0$ of $M^\dagger M$ is related to the pion mass, hence we argue that the step-size errors in the integration are amplified by some function of $1/m_\pi$. Wilson fermions do not have chiral symmetry hence $\lambda_0$ can wander arbitrarily close to zero during integration. On the other hand, for Staggered fermions $\lambda_0 > 0$ for all configurations with $m_q > 0$. Most likely the functional dependence on $1/m_\pi$ is different for Wilson and Staggered fermions, but in any case we expect very small pion masses and a small sea quark mass will be required for Staggered fermions before the magnified step-size errors will be an issue.

For our Wilson fermion tests, the first exercise we had to undertake was to find a suitable value for $\kappa$. We ran at $\kappa = 0.1680, 0.1677, 0.1675$, and $0.1670$: our results indicate that $m_\pi \lesssim 0.2/a$ for the first two values, indicating that $\kappa_c$ is probably in this region. If we extrapolate recently published results [3] we find that $\kappa_c \approx 0.1676$, in rough agreement with our data. For $\kappa = 0.1670$, $0.1675$ and $0.1677$ we find that $m_\pi \approx 0.47/a$, $0.33/a$ and $0.2/a$, respectively.

In Figure 1 we show the pion mass and acceptance rate as a function of MD time for our run at $\kappa = 0.1677$. Here, $R_{MD}$ is the accuracy of the normalized residual vector during integration. It is clear the system has a large exponential autocorrelation time and probably has only just reached equilibrium. It is also clear that as the correlation length grows the HMC acceptance rate falls, which requires reducing the integration step-size and/or the conjugate gradient accuracy in order to keep a reasonable acceptance rate. A preliminary study of the system indicates that this is primarily due to the increasing effect of high frequency fluctuations in the fermionic contributions to the action. In all our Wilson runs, we used a linear extrapolation of the proceeding two solutions for the initial guess of the conjugate gradient inverter.

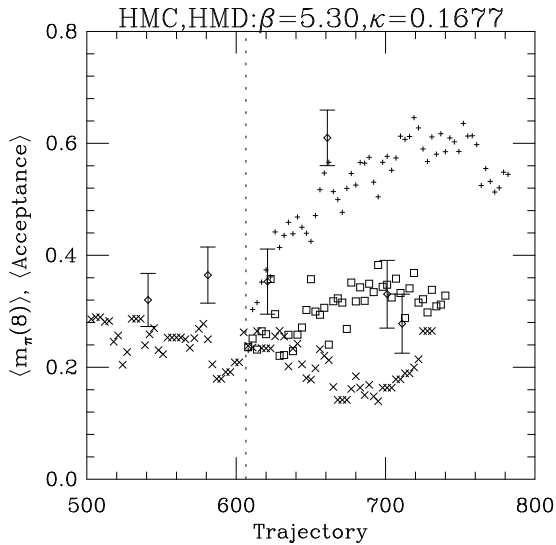

Figure 1. Time history of the effective pion mass (distance 8) for $\beta = 5.30$, $\kappa = 0.1677$ and $\delta\tau = 0.0069$ on a $16^3 \times 32$ lattice. Symbols are: ($\times$) HMC at $R_{MD} = 10^{-5}$ before trajectory 606 and $R_{MD} = 3 \times 10^{-7}$ after 606, ($+$) HMD at $R_{MD} = 10^{-5}$ and ($\square$) HMD at $R_{MD} = 3 \times 10^{-7}$. Also shown is the blocksize= 40 ($\diamond$) HMC acceptance uncorrected for autocorrelations.

We decided it would be useful to see what results we would get if we used the HMD algorithm instead of HMC for this system (i.e., if we omitted the Monte Carlo accept/reject step for $R_{MD} = 3 \times 10^{-7}$). It is commonly accepted that this should introduce small errors in physical quantities of order $\delta\tau^2$; to our surprise the HMD algorithm with the same parameters gave wrong answers for the pion mass. We also performed a HMD test using $R_{MD} = 10^{-5}$ and found the algorithm gave completely wrong answers for the pion mass. In fact other hints that the system contained light pions, such as the large number of CG iterations required per step to reach a given residual, also rapidly disappeared when running the HMD algorithm at $R_{MD} = 10^{-5}$. No shown is a test of when we reduced the step size to $\delta\tau = 0.004$ at $R_{MD} = 10^{-5}$, the HMD results were virtually identical to $\delta\tau = 0.0069$ indicating the $\delta H$ errors were dominated by the CG accuracy. We haven't performed any $\delta\tau$ HMD tests at $R_{MD} = 3 \times 10^{-7}$; however, we expect the CG accuracy to muddy the results. An even clearer indicator of $\delta H$ errors introduced is via the blocked spatial plaquette. At $\kappa = 0.1675$ and $0.1677$ and using $R_{MD} = 3 \times 10^{-7}$, the blocked plaquette for HMD deviated systematically from HMC when the acceptance test was turned off while the pion mass for $\kappa = 0.1675$ with HMD deviated only slightly from HMC.

These tests show that $\delta\tau$, CG accuracy and the acceptance are all very crucial for the algorithm to give correct physics. Tests of only step-size extrapolations are not complete enough since the CG accuracy is important for not just the minimization of $\delta H$ errors but for reversibility as well.

The parameters used in our Staggered simulations were based on the work of the MTc collaboration [4]. We used an $8^3 \times 16$ lattice at $\beta = 5.05$ and $m_q = 0.005$ and observed $m_\pi = 0.188(3)$ and $m_\rho \approx 1.0$. For HMC, we used $R_{MD} = 5 \times 10^{-7}$ and a linear extrapolation of the initial guess. To emphasize the $\delta H$ errors we used $R_{MD} = 5 \times 10^{-4}$ for HMD with a zero initial guess to maintain reversibility. We didn't find any statistically significant effects between HMD and HMC after 500 and 1000 trajectories each, respectively. The pion mass here is too small and the rho is too large to be realistic of the continuum limit, however. With this small lattice, we probably shouldn't expect the sea quark mass to be very small. We are currently running at the MTc parameters of $\beta = 5.35$ and $m_q \leq 0.01$ on a $16^3 \times 32$ lattice which we hope will provide a test more useful for calculations on future machines.

In a separate study, we carried out a more accu-rate MD integration with Wilson fermions using the higher-order integration schemes [6], with the hope of finding a cheap way of increasing the acceptance rate for HMC. Theoretical analyses of these algorithms were based upon the idea that because the change in energy over a trajectory would be $\delta H = O(\delta\tau^4)$ instead of $\delta H = O(\delta\tau^2)$ one could use a much larger step size to get the same acceptance rate. This does not work in practice: while the Campostrini scheme does indeed give much smaller values for $\delta H$, the integration becomes unstable if any attempt is made to increase $\delta\tau$. Indeed, the Campostrini method proved unstable unless the step size was reduced slightly from that used in our HMC runs. This is, of course, the behavior we should have expected. The limit on the step size is given by the constraint from stability ($\omega\delta\tau \ll 1$ where $\omega$ is the highest frequency of the physical system) and not by the $\text{erfc}(const \cdot \delta\tau^2\sqrt{V})$ dependence of the acceptance rate on the volume (at least for the volumes we have used).

Initially we thought the extra computation required for the Campostrini method was small because the interpolated solutions of the previous CG solutions were very accurate. This was erroneous, as we had overlooked the fact that the delicate cancellations required by the Campostrini method required a much more accurate measurement of $H$ for each step, and thus a much lower CG residual. When we lowered the CG residual to the value required to make the CG solution error contribution to $H$ smaller than the true step-size errors, we found that the magnitude of $\delta H$ was greatly reduced but at the cost of many more CG iterations. Sadly we found that unless enough extra CG iterations were performed the use of interpolated initial guesses for the CG solver induced an irreversibility in the HMC algorithm which led to clearly incorrect results.

Our conclusions on this issue are that one must either make sure that the CG residual is small enough to avoid such systematic biases or one must use a time-symmetric initial guess for the CG solution such as zero. Both options make the Campostrini method much more expensive than simple leapfrog integration. Which method is ultimately cheaper is still under investigation.

If we had carried out an HMD calculation with similar $\delta H$ errors used in the present HMC computations, we would have been forced to use a larger value of $\kappa$ in order to measure a light pion mass. This mass would have been much smaller than the inverse correlation length actually present in the configurations. Indeed, we may well have been measuring valence observables on an almost-quenched system, so we would have concluded that physics with light dynamical quarks was very similar to quenched physics with the same masses. We would also have found a much smaller relaxation time and autocorrelation time, and thus been misled into thinking that full QCD calculation were much cheaper than they really are.

What are the implications of this for staggered quarks calculations done using the HMD algorithm? While our Wilson quark results do not tell us directly about the behavior of staggered quarks — the effects we see might possibly just be artifacts of Wilson quark dynamics — they lead us to suggest that it is necessary for a careful zero step-size extrapolation to be done for any HMD calculation, with special attention required to verify that the system is truly in equilibrium.

## REFERENCES

1  A.D. Kennedy, Intl. J. Mod. Phys. C3, 1 (1992); J.C. Sexton and D.H. Weingarten, Nucl. Phys. B30 (1992) 665.
2  S. Gottlieb,W. Liu,D. Toussaint,R.L. Renken, R.L. Sugar, Phys. Rev. D35 (1987) 2631.
3  R. Gupta,C. Baillie,R. Brickner,G. Kilcup,A. Patel and S. Sharpe, Phys. Rev. D44 3272 (1992).
4  E. Laermann, R. Altmeyer, K.D. Born, W. Ibes, R. Sommer, T.F. Walsh and P.M. Zerwas, Nucl. Phys. B (Proc. Suppl.) 17 (1990) 436, Lattice 89.
5  A.D. Kennedy and B. Pendelton, Nuc. Phys. B20 (Proc. Suppl.) 118 (1991); S. Gupta, Phys. Let. B278, 317 (1992); Intl. J. Mod. Phys. C3, 43 (1992).
6  M. Campostrini and P. Rossi, Nuc. Phys. B329, 753 (1990); M. Creutz and A. Gocksch, Phys. Rev. Let. 63, 9 (1989).